\documentclass[prl,aps,longbibliography,twocolumn,superscriptaddress]{revtex4}

\usepackage{graphicx}
\usepackage{times}
\usepackage{color}
\usepackage{float}
\usepackage{epsfig}
\usepackage{epstopdf}

\begin{document}

\title{The {\it Bellerophon} state: a novel coherent phase of globally coupled oscillators}

\author{Hongjie Bi}\thanks{Authors contribute equally to this work.}
\affiliation{Department of Physics, East China Normal University, Shanghai 200241, China}

\author{Xin Hu}\thanks{Authors contribute equally to this work.}
\affiliation{Key Lab of Nanodevices and Applications-CAS \& Collaborative Innovation Center of Suzhou Nano Science and Technology, Suzhou Institute of Nano-Tech and Nano-Bionics, Chinese Academy of Sciences, Suzhou 215123, China. }

\author{S. Boccaletti}\thanks{Corresponding author: stefano.boccaletti@gmail.com}
\affiliation{CNR-Institute of Complex Systems, Via Madonna del Piano, 10, 50019 Sesto Fiorentino, Florence, Italy}
\affiliation{The Embassy of Italy in Tel Aviv, 25 Hamered street, 68125 Tel Aviv, Israel}

\author{Xingang Wang}
\affiliation{School of Physics and Information Technology, Shaanxi Normal University, Xi¡¯an 710062, China}

\author{Yong Zou}
\affiliation{Department of Physics, East China Normal University, Shanghai 200241, China}

\author{Zonghua Liu}
\affiliation{Department of Physics, East China Normal University, Shanghai 200241, China}

\author{Shuguang Guan}\thanks{Corresponding author: guanshuguang@hotmail.com}
\affiliation{Department of Physics, East China Normal University, Shanghai 200241, China}

\begin{abstract}
From  rhythmic physiological processes to the collective behaviors of technological and natural networks, coherent phases of interacting oscillators are the foundation of the events' coordination leading a system to behave cooperatively. We unveil the existence of a new of such states, occurring in globally coupled nonidentical oscillators in the proximity of the point where the transition from the system's incoherent to coherent phase converts from explosive to continuous. In such a state, oscillators form quantized clusters, where they are neither phase- nor frequency-locked. Oscillators'  instantaneous speeds are different within the clusters, but they form a characteristic cusped pattern and, more importantly, they behave periodically in time so that their average values are the same. Given its intrinsic specular nature with respect to the recently introduced Chimera states, the phase is termed  the {\it Bellerophon} state. We provide analytical and numerical description of the microscopic and macroscopic details of {\it Bellerophon} states, thus furnishing practical hints on how to seek for the new phase in a variety of experimental and natural systems.

\end{abstract}

\maketitle

The emergence of coherent phases of interacting oscillators is one of the most important phenomena in nature, as it represents the foundation for the cooperative functioning of a wealth of different systems. To gather insights and understanding on the mechanisms and reasons underlying this organizational behavior, physicists resort on solvable and simplified frameworks, as the Kuramoto \cite{kuramoto} and Kuramoto-like \cite{Zhang2013,Hu2014,Zhang2015} models, where a variety of collective states can be described: from full \cite{Strogatz1991,Strogatz2000} to cluster \cite{cluster}, to explosive synchronization (ES) \cite{Zhang2015,Gardenes2011}. Recently, special interest has been devoted to Chimera states (CS) \cite{kura1,abrams}, which consists in a transient \cite{ocho} coexistence of coherent and incoherent domains in locally coupled identical oscillators. Various types of Chimera states, such as the breathing CS \cite{abrams2}, the clustered CS \cite{sethia} and the multi-CS \cite{omelchenko} have been theoretically described, and observed in experiments \cite{experi}.

In this Letter we give evidence of a previously unknown coherent phase in globally coupled oscillators. The novel phase is proper of nonidentical oscillators with widely different frequencies, and emerges as an asymptotic state in the proximity of a tri-critical point, where the transition from the system's incoherent to coherent behavior converts from explosive to continuous. There, oscillators form quantized clusters, where they are neither phase- nor frequency-locked. Rather, each of the oscillators'  instantaneous speeds is different within the clusters, but the instantaneous frequencies form the same cusped pattern characterizing the average speeds of CS. The oscillators's instantaneous frequencies behave periodically in time so that their average values are the same. Because of its intrinsic specular nature with respect to CS, the new phase is termed here  the {\it Bellerophon} state, as Bellerophon was the name of the great hero who, in Greek mythology, confronted with (and eventually killed) the monster Chimera.

\begin{figure}[h]
\begin{center}
\includegraphics[width=0.5\textwidth]{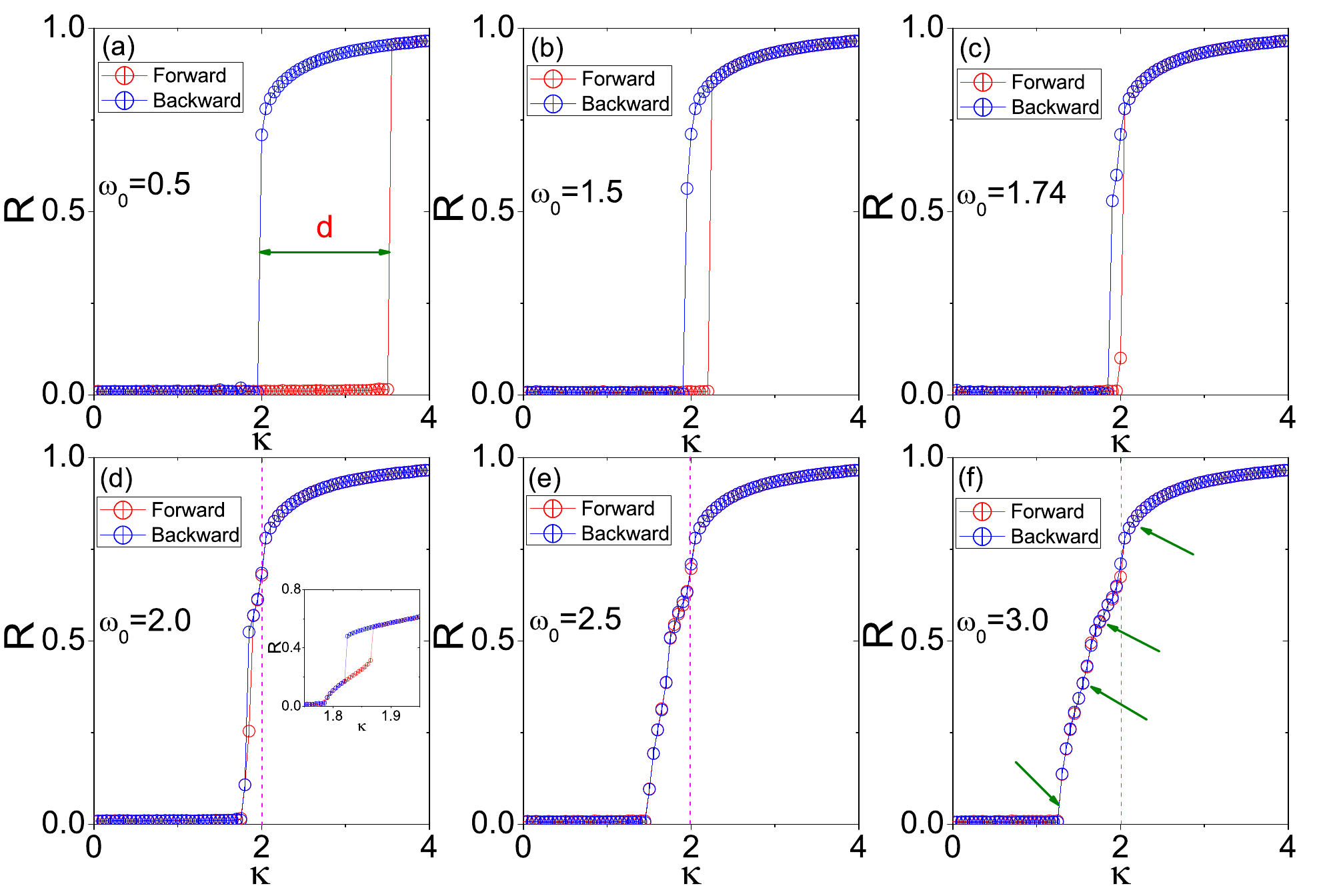}
\caption{(Color online) {\bf From explosive to continuous transition.} $R$  {\it vs}. $\kappa$ for model (\ref{eq:model}).
$\Delta =1$ and $\omega_0=0.5$ (a), 1.5 (b), 1.74 (c), 2.0 (d), 2.5 (e), and $3.0$ (f).}\label{fig-R}
\end{center}
\end{figure}

Let us start by considering a Kuramoto-like model of $N$ globally coupled phase oscillators. The model reads:
\begin{equation}\label{eq:model}
\dot{\theta}_i=\omega_i + \frac{\kappa | \omega_i |}{N}\sum_{j=1}^{N}\sin(\theta_j-\theta_i),\quad i=1,...,N,
\end{equation}
where $\theta_i$ and $\omega_i$  are the instantaneous phase and the natural frequency of the $i$th oscillator, respectively.  Dots denote temporal derivatives, and $\kappa$ is the coupling strength. The level of synchronization is measured by the order parameter $R=\frac{1}{N}\langle\vert\sum_{j=1}^Ne^{i\theta_j}\vert\rangle_T$, where $\vert \cdot \vert$ and $\langle \cdot \rangle_T$ denote module and time average, respectively.   The set of natural frequencies $\{\omega_i\}$ is drawn from a given frequency distribution (FD) $g(\omega)$, which
is assumed to be an even function [$g(\omega)=g(-\omega)$], symmetric, and centered at 0. We take $g(\omega)=\frac{\Delta}{2\pi} [\frac{1}{(\omega-\omega_0) ^2 + \Delta ^2} +\frac{1}{(\omega+\omega_0) ^2 + \Delta ^2  }],$ to be a bimodal Lorentzian distribution
with $\pm \omega_0$ being the location of the two peaks,
and $\Delta$ the width at half maximum of the FD. Notice that, depending on $\omega_0/\Delta$, such a FD can be, in fact, either uni-modal ($\omega_0/\Delta\leq\sqrt{3}/3$) or bimodal ($\omega_0/\Delta>\sqrt{3}/3$).

As a first step, we investigate numerically the synchronization transition in system (\ref{eq:model}).
As the considered FD has two basic parameters ($\omega_0$ and $\Delta$),  we fix for simplicity $\Delta=1$ and let
$\omega_0$ to increase (which, physically, is tantamount to progressively increasing the distance between the two peaks of the FD).
The results are shown in Fig. \ref{fig-R} \cite{notilla}. For small values of $\omega_0$, an irreversible, first-order-like, abrupt transition [Fig. \ref{fig-R}(a)] is observed, featuring a characteristic hysteresis area whose width can be defined as $d=\kappa_b-\kappa_f$ (with $\kappa_b$ and $\kappa_f$ being the critical points for the backward and forward transitions, respectively).
Figs. \ref{fig-R}(b), (c) show that the width of hysteresis progressively shrinks as $\omega_0$ increases.
Eventually, when $\omega_0$ is large enough [Figs. \ref{fig-R}(d), (e), (f)], one observes a reversible, second-order-like, continuous transition.
The results then show that system (\ref{eq:model}) sustains both a first- and a second-order-like transition to synchronization.

\begin{figure}[h]
\begin{center}
\includegraphics[width= 0.35\textwidth]{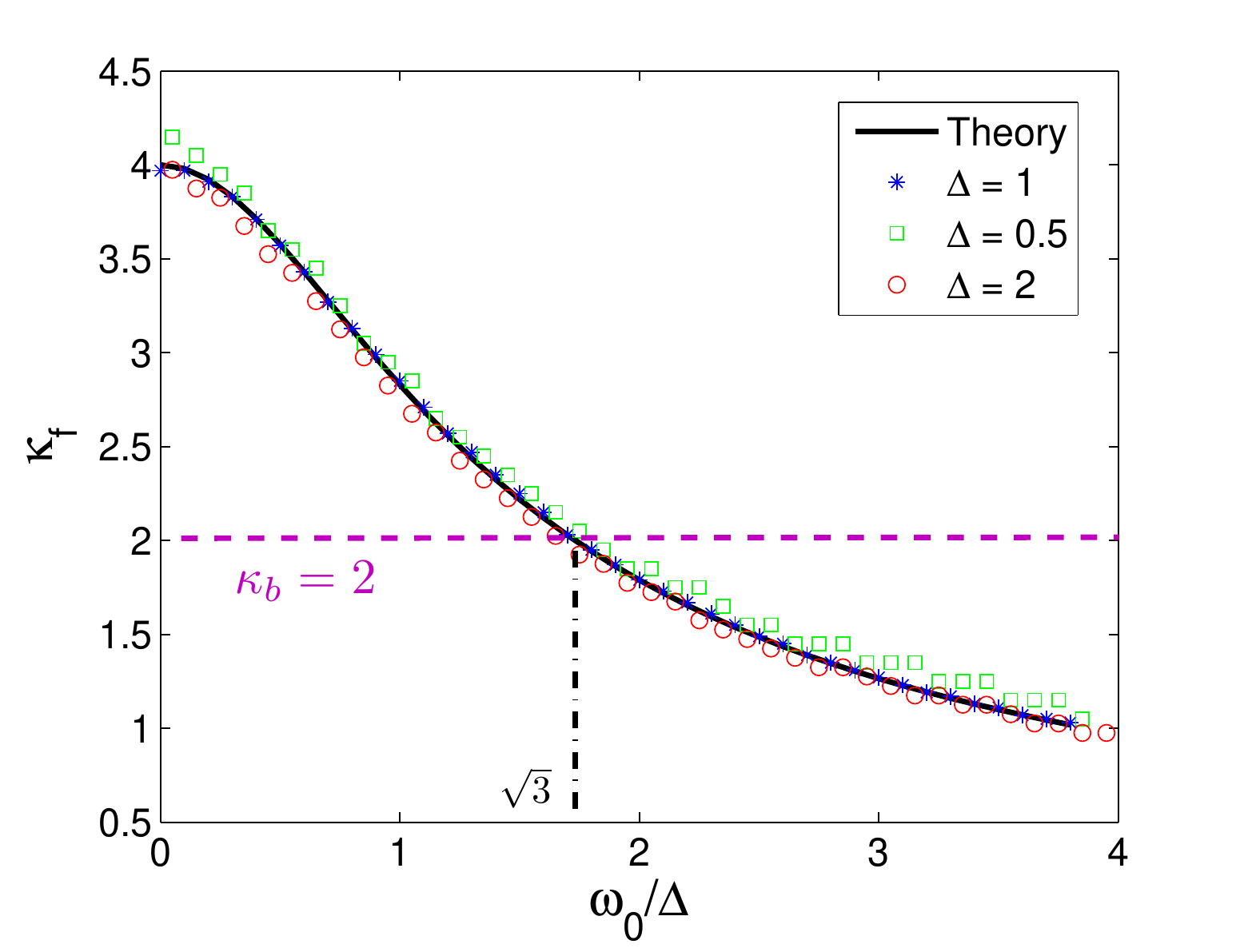}
\caption{(Color online). {\bf Critical point for the forward transition.}
$\kappa_f$ vs. $\omega_0 / \Delta$. The black curve corresponds to the analytic solution (\ref{eq:kf}). The purple dashed line marks the
backward transition point. The theoretical prediction and the numerical results coincide perfectly. }\label{fig-kf}
\end{center}
\end{figure}

\noindent
As long as the FD is symmetric (as in the present case), the critical point for the backward transition is $\kappa_b =2$ \cite{Hu2014}, which is fully verified in our simulations [see panels (a)-(c) of Fig. \ref{fig-R}]. However, the critical point $\kappa_f$ for the forward transition actually varies with $\omega_0$ (at $\Delta=1$), thus inducing the hysteresis area to shrink, and leading eventually to the observed conversion from an ES to a continuous transition to synchronization (occurring at $\omega_0 \approx 1.7$).
The second step of our study is then seeking for an analytic solution for $\kappa_f$, which turns out to be (full details contained in the {\it Supplementary Material}):
\begin{equation}\label{eq:kf}
\kappa_f=\frac{4}{\sqrt{1+(\omega_0 / \Delta)^2}}.
\end{equation}
\noindent
Eq. (\ref{eq:kf}) tells that the critical point  for the forward transition is uniquely determined by the dimensionless parameter $\omega_0/\Delta$. In particular, when $\omega_0=0$, the FD degenerates into the typical unimodal Lorentzian distribution, and
Eq. (\ref{eq:kf}) predicts $\kappa_f=4$, in full agreement with what reported in Ref. \cite{Hu2014}. Fig. \ref{fig-kf} gives account that Eq. (\ref{eq:kf}) is remarkably well verified by numerical simulations (at all values of $\Delta$ and within the entire range of $\omega_0/\Delta$).

\begin{figure}[h]
\begin{center}
\includegraphics[width= 0.48\textwidth]{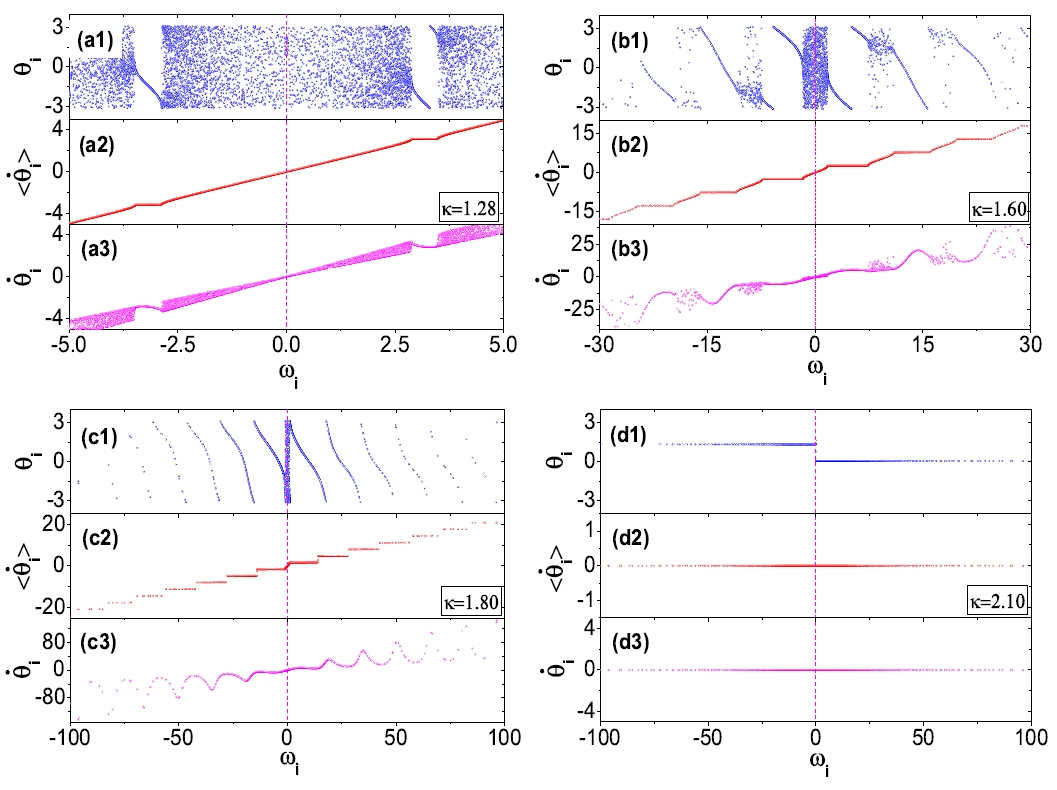}
\caption{(Color online). {\bf  The {\it Bellerophon} states.} Snapshots of the instantaneous phase $\theta_i$ (upper plots), the instantaneous speed $\dot{\theta}_i$ (middle plots), and the average speed $\langle \dot{\theta}_i \rangle$ (lower plots) {\it vs.} natural frequencies $\{\omega_i\}$ of the oscillators. $\kappa=1.28$ (a), 1.60 (b), 1.80 (c), and $2.10$ (d, the fully synchronized state). All other parameters specified in the text.
Panels (a)-(c) refer to the {\it Bellerophon} states.
 }\label{fig-phase}
\end{center}
\end{figure}

\noindent
Furthermore, Eq. (\ref{eq:kf}) fully explains the conversion from the first- to the second-order-like transition observed in Fig. (\ref{fig-R}). As it is seen in Fig. (\ref{fig-kf}), $\kappa_f$ decreases monotonically as $\omega_0/\Delta$ increases, causing (as $\kappa_b=2$ always) the hysteresis area to shrink monotonically. When  $\omega_0/\Delta=\sqrt{3}$, $\kappa_f=\kappa_b=2$,  and the forward and backward transition points almost coincide
[see Fig. \ref{fig-R}(c)]. As $\omega_0/\Delta$ gradually exceeds $\sqrt{3}$,
the hysteresis area does not immediately disappear [see the inset of Fig. \ref{fig-R}(d)]. Actually, at $\omega_0/\Delta=\sqrt{3}$, a Hopf bifurcation occurs during both the forward and backward processes, and both bifurcations are continuous. For the forward direction, the system first undergoes a continuous transition, followed by an ES transition (as $\kappa$ further increases). A similar scenario of transitions characterizes also the backward direction.
An initial parameter regime $\omega_0/\Delta>\sqrt{3}$ then exists, where the system undergoes the cascade of one continuous and one explosive transition during both forward and backward continuity. A further increase of $\omega_0/\Delta$ causes the hysteresis area to eventually disappear [Figs. \ref{fig-R}(e) and (f)], leading to a situation where only continuous transitions occur in the system. It is in this latter regime, i.e. close to a tri-critical point in parameter space that novel coherent phases, the {\it Bellerophon} states, emerge in the path leading the system from its unsynchronized to its synchronized behavior.

\begin{figure}[h]
\begin{center}
\includegraphics[width= 0.48\textwidth]{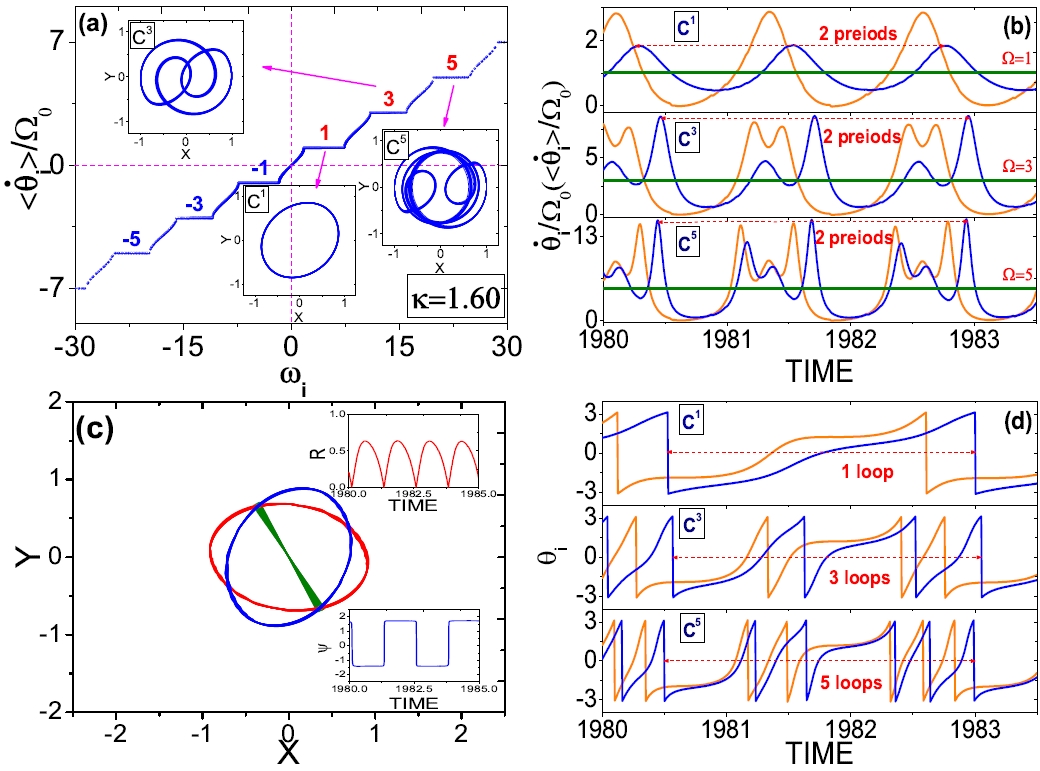}
\caption{(Color online). {\bf The state of Fig. \ref{fig-phase}(b)}.
(a) The staircases of average speeds for coherent clusters. It is evident how they correspond to odd-numbered multiples of the principle frequency $\Omega_1$.
(b) Time series of the instantaneous speeds of clustered oscillators. In the panel, two sample oscillators are arbitrarily chosen for  clusters $C^1$ (top), $C^3$ (middle), and $C^5$ (bottom). Straight lines mark the average speed.
(c) Order parameters for all oscillators with positive (blue oval) and negative (red oval) frequency, and order parameter for
all oscillators (green lines). The insets are the time series of the global order parameter $R(t)$ and $\Psi(t)$, which are typically oscillatory. (d) Time series of the instantaneous phases corresponding to (b).
 }\label{fig-1.6}
\end{center}
\end{figure}

\noindent
The third step of our study is to provide a full characterization of such a novel state, and to discuss the relevant differences with other typical coherent states of Kuramoto-type models.
For the sake of exemplification, we take the case of $\omega_0/\Delta=3$ [Fig. \ref{fig-R}(f)]. Here, the system exhibits two continuous transitions at $\kappa_1=4/\sqrt{10}\approx1.26$ and $\kappa_2=2$, respectively. Therefore, three parameter regimes can be identified: $\kappa<\kappa_1$ (I),  $\kappa_1<\kappa<\kappa_2$ (II), and $\kappa>\kappa_2$ (III). In regime I, the coupling strength is small, and the system features the trivial incoherent state. In regime III, the coupling is so strong that the system goes into the fully synchronized state, in which all oscillators split into two fully synchronized clusters. {\it Bellerophon} phases are steady states occurring in the middle regime II, i.e. during the path to full synchronization.
In Fig. \ref{fig-phase} four typical phases are illustrated, corresponding to the $\kappa$ values denoted by letters A, B, C, and D in Fig. \ref{fig-R}(f). They are characterized by three quantities: the instantaneous phases $\theta_i$, the instantaneous (angular) speed $\dot{\theta}_i$, and the average speed $\langle \dot{\theta}_i\rangle$ (i.e., the oscillators' effective frequencies), where the bracket stands for a long time average.

\begin{figure}[h]
\begin{center}
\includegraphics[width= 0.48\textwidth]{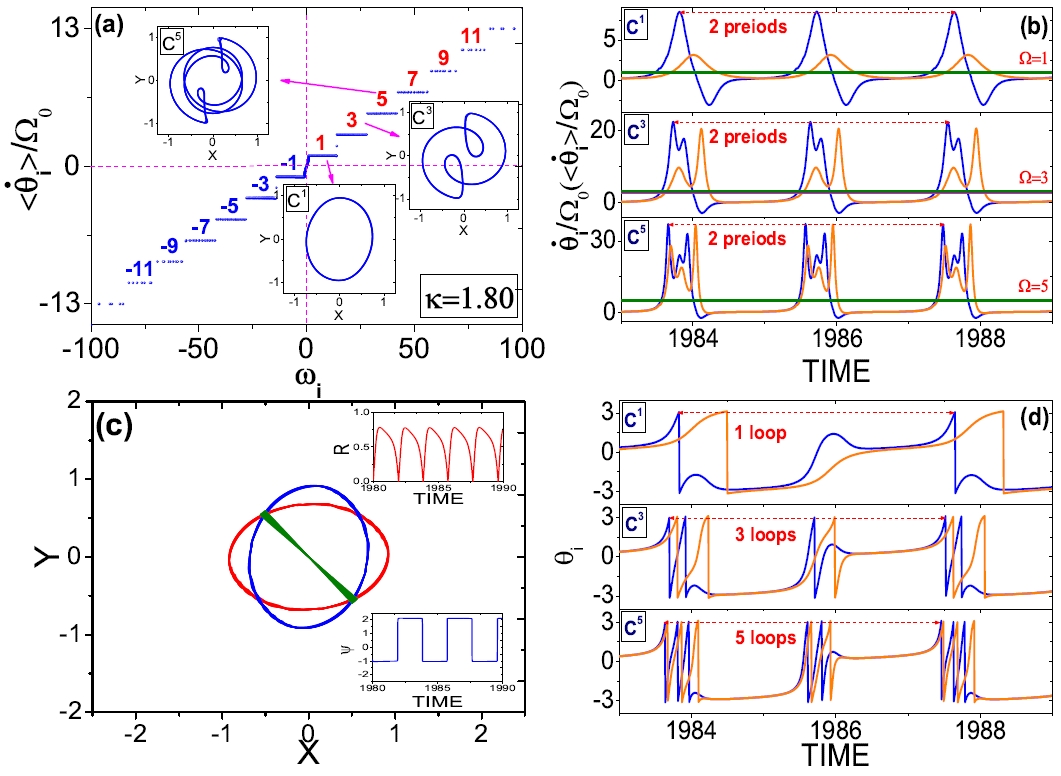}
\caption{(Color online) {\bf The state of Fig. \ref{fig-phase}(c).}
Same quantities reported as in the caption of Fig. \ref{fig-1.6}.
In panel (b), negative instantaneous speeds are observed during certain time intervals, and the phases decrease accordingly
as in (d). This implies that although some oscillators go positively on average, they do go back for certain time intervals during one loop, just like shuttle-run.
 }\label{fig-1.8}
\end{center}
\end{figure}

In Fig. \ref{fig-phase}(a), $\kappa=1.28$. As $\kappa$ just exceeds $\kappa_1=1.26$, two small symmetric clusters emerge, whose average speeds are equal to each other in magnitude, but opposite in sign. Oscillators in the two clusters rotate with the same average speed (but different instantaneous phases and frequencies). At $\kappa=1.60$ [Fig. \ref{fig-phase}(b)], a multi-clustered state emerges. The number of clusters increases {\it in pairs} as $\kappa$ increases, each pair containing oscillators which are symmetric in terms of their natural frequencies. Oscillators inside each cluster have the same average speed [see the staircase structure of Fig. \ref{fig-phase}(b2)],  but different instantaneous frequencies [Fig. \ref{fig-phase}(b3)]. The clusters coexist with drifting oscillators that are not synchronized.
In Fig. \ref{fig-phase}(c), $\kappa=1.80$. This is also a {\it Bellerophon} state, but different from that of Fig. \ref{fig-phase}(b).  The coherent clusters now occupy almost all the range of natural frequencies, except for a small narrow zone around the central frequency, the increase of $\kappa$
results in all drifting oscillators being gradually recruited into either one of the clusters.
Finally, Fig. \ref{fig-phase}(d) ($\kappa=2.10$) refers to the fully coherent phase, where two giant clusters are formed. In each cluster, oscillators with positive or negative frequencies coincide with each other totally: they feature now the same instantaneous speed, so that the whole system behaves like two giant oscillators.

A much better insight is gathered by inspecting the system's macroscopic and microscopic details. Fig. \ref{fig-1.6}(a) reveals that the staircases of coherent clusters at $\kappa=1.60$ satisfy in fact a certain rule: they are {\it quantized}, and can be expressed as $\pm(2n-1)\Omega_1, n= 1, 2, \cdots$ \cite{Engelbrecht2012}, where $\Omega_1$ is the lowest frequency, i.e., the principle (or base) system's frequency.  Accordingly, depending on their multiple to $\Omega_1$, clusters can be named as $C^1, C^3, C^5, \cdots$, respectively. The key, and also subtle, point here is that,
although the average speeds of oscillators inside each cluster are equal to each other, their instantaneous speeds are generally different and quite heterogeneous, forming the characteristic cusped pattern totally analogous to that featured by the average frequencies of the oscillators within the CS. Fig. \ref{fig-1.6}(b) shows that the instantaneous speeds of oscillators inside the same cluster evolve periodically, but
different oscillators follow different periodic patterns. In other words, the instantaneous speed for each oscillator evolves uniquely.
This makes  {\it Bellerophon} states essentially different with respect to other coherent states observed in Kuramoto-like models, such as the partially coherent phase \cite{Strogatz2000}, the standing wave \cite{Martens2009,Pazo2009}, the traveling wave states \cite{Martens2009,Iatsenko2013}, and the CS \cite{kura1,abrams},
 where oscillators inside the coherent cluster are typically frequency-locked.
Moreover, even though the instantaneous speed of the clusters' oscillators varies non-uniformly during one period (particularly for those clusters with large $n$), the average speeds during one period for all oscillators in a certain cluster turn out to be the same, i.e., an odd-numbered multiple of $\Omega_1$.
As the instantaneous speed characterizes the rotations of oscillators along the unit circle, very interesting collective motion of oscillators is observed [Fig. \ref{fig-1.6}(d)]: during one period $T_1=1/\Omega_1$, the oscillators in $C^1$ all perform one loop along the unit circle, and in the mean time,  the oscillators in $C^3$ and $C^5$ rotate three loops and five loops, respectively. In analogy, oscillators in $C^{2n-1}$ will perform $2n-1$ loops.
Compared with Fig. \ref{fig-1.6}(b), we further find that during one loop, the instantaneous speeds for all coherent oscillators experiences two periods, i.e., each oscillator repeats its motion  during the two half periods.
In Fig.  \ref{fig-1.6}(a), we report the local value of the order parameter (i.e.that contributed by only those oscillators in a certain cluster) in  the complex plane, for $C^1$, $C^3$ and $C^5$. Due to the complicated phase relationships among oscillators in each cluster [see Fig. \ref{fig-1.6}(d)], the resulting value is typically periodic or quasi-periodic, and follows a complicated orbit. Essentially, each cluster can be seen as a giant oscillator, with properties described by the local order parameter. Fig. \ref{fig-1.6}(c) reports the order parameters for all positive and negative frequencies (including the drifting oscillators). In phase space, orbits appear as two smeared ovals, reflecting the quasi-periodic motion of the total order parameter, as shown in the insets of Fig. \ref{fig-1.6}(c).

Similarly, Fig. \ref{fig-1.8} characterizes the {\it Bellerophon} state at $\kappa=1.80$. The emerging multi-clustered phase shares common features with that of Fig. \ref{fig-1.6}, together with other new properties. First [Fig. \ref{fig-1.8}(a)], all oscillators join here the synchronized  clusters, except for a small fraction localized in a narrow zone of frequencies around 0. Second [Fig. \ref{fig-1.8}(b)], although the average speed is positive, some oscillators have a negative instantaneous speed within two time intervals in one period, meaning that those units rotate inversely twice during one loop.
Fig. \ref{fig-1.8}(d) shows the behavior of two time intervals in one period, during which the phases of those oscillators decrease. Third, a careful examination reveals that the principle  period  $T_1$ becomes larger and larger as $\kappa$ increases (with  $T_1 \to \infty$ as $\kappa \to 2$, where the system enters a fully synchronized state of two static clusters) [Fig. \ref{fig-phase}(d)].
In summary, the system (for moderately large values of $\kappa$) is made of multi-cluster oscillatory coherent states.
It should be remarked that, besides the situations of Figs. \ref{fig-R}(e) and (f),
such oscillatory states occur also during ES, as for instance, in Fig. \ref{fig-R}(d) ($1.7<\kappa<2$).
In these states, oscillators form symmetric synchronized cluster pairs, with quantized average speeds.  Oscillators having the same average speed have different instantaneous speeds, forming a characteristic cusped pattern and leading to very complicated (yet collectively organized) motions along the unit circle.
For a better understanding of the overall picture characterizing {\it Bellerophon} states, the {\it Supplementary Material} contains several animated movies which help visualizing the evolution of phases, speeds, and motions of oscillators on the unit circle.

In conclusion, we have given evidence of a novel, asymptotic, coherent phase of globally coupled oscillators: the {\it Bellerophon} state. Such a state occurs for nonidentical oscillators, where frequencies are widely distributed, and in proximity of a tri-critical point in parameter space. As so, it is essentially different from all previous phases described in Kuramoto-like models.
Within the novel state, oscillators form quantized clusters, where they are neither phase- nor frequency-locked. Oscillators'  instantaneous speeds are different, but they behave periodically, and most importantly, their average values are the same.
{\it Bellerophon} states represent then a higher-order form of coherence:  individual oscillators inside each cluster do effectively influence each other,
at the same time at which they still have a relative freedom.
While revealing functional relationships within each cluster will certainly be a mathematical challenge for the future, our analytical and numerical description will certainly help physicists to seek {\it Bellerophon} states in a variety of experimental and natural systems.

Work partly supported by the National Natural Science Foundation of China under the Grants  No. 11305062, 11375066, 11135001, and 11375109;
and
the Fundamental Research Funds for the Central Universities under the Grant No. GK201303002.

\bibliographystyle{apsrev}

\begin{thebibliography}{99}
\bibitem{notilla} Numerical values are obtained by a fourth order Runge-Kutta integration method, with integration time step 0.005. $N=10,000$.
\end{thebibliography}

\end{document}